\documentclass[preprint2]{aastex}

\begin{document}

\title{Direct determination  of the epicycle frequency in the galactic disk,
 and the derived rotation velocity $V_0$}
\author{ J.R.D. L\'epine\footnote{ Instituto de Astronomia, Geof\'isica e 
Ci\^encias Atmosf\'ericas, Universidade
 de S\~ao Paulo, Cidade Universit\'aria, S\~ao Paulo, SP, Brazil; 
E-mail: jacques@astro.iag.usp.br}
\and
Wilton S. Dias \footnote{UNIFEI, Instituto de Ci\^encias Exatas, Universidade
Federal de Itajub\'a, Itajub\'a, MG, Brazil 
}
\and Yuri Mishurov \footnote {South Federal University (Rostov State
University), Rostov-on-Don,  Russia}
 }

\begin {abstract} 

We present a method which allows a direct measurement of the epicycle frequency
$\kappa$ in the galactic disk, using the large database on open clusters completed by our
group. The observed velocity vector (amplitude and direction) of the clusters in
the galactic plane is derived from the catalog data.   In the epicycle approximation,
this velocity is the sum of the circular velocity, described by the galactic rotation
curve, and of a residual velocity, which has a direction that  rotates with the
frequency $\kappa$.  If for some reason the clusters are formed with non-random initial
perturbation velocity direction (measured for instance with respect to the direction of
circular rotation), then a plot of the orientation angle of the residual velocity as
a function of age reveals the epicycle frequency.  The data analysis confirms that
this is the case; due to the non-random initial velocities, it is possible to measure 
$\kappa$ for different galactic radii. Our analysis considers that the effect of the
arms on the stellar orbits is small (the Galactic potential is mainly axis-symmetric); 
in this sense our results do not depend on any specific model of the spiral structure,  
like the existence of a given number of spiral arms, or on a particular choice of the
radius of corotation.  The values of $\kappa$ provide constraints on the rotation 
velocity of the disk and on its minimum beyond the solar radius; in particular, 
 $V_0$ is found to be 226 $\pm$ 15 kms$^{-1}$ even if the short scale (R$_0$ = 7.5 kpc)
 of the galaxy is adopted. The mesured $\kappa$ at the solar radius is 42$\pm$4 
 kms$^{-1}$kpc$^{-1}$.

\end{abstract}


% ------------------------ Section 1 ---------------------

\section{Introduction}

Open clusters are ideal objects to study the dynamics of the spiral structure of 
the Galaxy. A large number of them have known distances and space velocities 
(proper motions and radial velocities), that are more precisely determined than
those of individual stars. Another important parameter is their age, which is
obtained by means of the HR diagrams. In a previous work, Dias \& L\'epine (2005,
hereafter DL) integrated backwards the orbits of a sample of open clusters to 
find their birthplaces as a function of time, and obtained the velocity of the
spiral arms. In a similar manner, by integrating the orbits, it is possible to
determine the velocity (direction and amplitude) of the clusters at the time
of their birth. Since the gas of the disk is known to present systematic 
perturbations with respect to circular rotation, or "streaming motions" related
to the spiral structure, it would not be surprising if the new born clusters also
present some systematic deviations from circular motion. The study of initial 
velocities allows us to verify if the dominant process of star formation produces
clusters with random velocities or, on the contrary, with velocities oriented in 
some preferential direction(s). The answer to this question will certainly be 
able to put restrictions to the mechanisms that trigger star formation.

One result of our investigation is the discovery that part of the open
clusters have initial velocities in a constant narrow angle for long period of
times. This property allowed us to develop a method for a direct determination
of the epicycle frequency. According to the perturbation 
theory, if an open cluster was born with an initial velocity different from that
of the circular velocity given by the rotation curve, it will follow an orbit that
can be described by the epicycle approximation.  The motion can be seen as the sum
of the circular motion around the galactic center, plus a small amplitude rotation
around the equilibrium point in the circular orbit. The epicycle frequency $\kappa$
is a function of the galactic rotation velocity $\Omega$ and of its derivative
  $d\Omega/dR$ only:
 $$\kappa^2 = 4 \Omega^2 (1 + \frac{1}{2} \frac{R}{\Omega} \frac{d\Omega}{dR}) ~~~~   (1)$$
Based on this equation, the  epicycle frequency is easily derived from any observed
rotation curve. However, since there are uncertainties  concerning the exact shape of the
rotation curve, as several different curves are proposed in the literature, a direct
measurement of the epicycle frequency, without passing through equation (1), is 
of great interest. Such a measurement, in turn, can help to select the best galactic
parameters and rotation curve. We discuss in this work, in addition to the statistics
of initial velocities of open clusters,  the restrictions on the
galactic parameters that can be derived from the measured epicycle frequency. 

\section{The Open Clusters Catalog}

We make use of the {\it New Catalogue of Optically visible Open Clusters and Candidates} 
published by Dias et al. (2002) and updated by Dias et al. (2006) \footnote{Available
at the web page http://astro.iag.usp.br/$_{\symbol{126}}$wilton}. This catalog updates the previous
catalogs of Lyng\"a (1987) and of Mermilliod (1995). The present 2.7 version of the catalog
contains 1756 objects, of which 850 have published ages and distances, 889 have
published proper motions (most of them determined by our group, Dias et al. 2006 and 
references therein) and 359 have radial velocities. Recently, Pauzen and Nepotil (2006)
compared statistically our catalog with averaged data from the literature 
and reached a result which is in favor of the data from the catalog.

\section{The method}

The main result  of  the perturbation theory of stellar orbits in a central potential,
applied to the galactic disk, is that the motion of a star (or of an open cluster)
can be described as the sum of a circular rotation around the galactic center, 
with angular velocity $\Omega(R)$,  plus an epicycle perturbation, as illustrated in 
Figure 1 ( eg. Binney \& Tremaine, 1987, hereafter BT87).
The epicycle perturbation is an oscillation with frequency $\kappa$ around the
unperturbed circular orbit, in both the radial direction and the direction of circular
rotation: 
$$r(t) = r_0 + \xi(t), ~~~         with ~ \xi(t)  = b ~sin (\kappa t + \phi)~~~(2)$$
$$\theta(t) = \theta_0  + \Omega t   +  \frac{\eta (t)}{r_0}, ~~~   with ~ \eta(t) =  
a ~ cos(\kappa t + \phi)~~~(3)$$
($\xi$ and $\eta$ represent the displacements in the two directions,  $b$ and $a$
the amplitudes in units of length, and $\phi$ the phase). In the frame of reference 
of the guiding center of the epicycle motion, 
rotating with angular velocity $\Omega$, the perturbed displacement describes an 
ellipse, with the  ratio of the amplitudes $ a/b  =   2 \Omega/\kappa$  (BT87), 
which is approximately 1.5. Like in the case of a circular motion, the perturbed
velocity rotates with the same frequency of the perturbed  displacement.  The 
velocity components are:
$$\xi'=\frac{d\xi}{dt} = b~\kappa~cos (\kappa t + \phi)~~~(4) $$
$$\eta'= \frac{d\eta}{dt} =  -a ~\kappa~sin(\kappa t + \phi)~~~(5)$$
The ratio of the velocity amplitudes is the same of the displacement amplitudes. 
The angle $\gamma$ between the perturbed velocity vector and the direction of circular 
rotation, taken as a reference (see Figure 2), is: 
$$ \gamma(t) = tan^{-1} (\frac{\xi'}{\eta'}) = tan^{-1} [(b/a) tan(\kappa t + \varphi)]~(6)$$ 
We introduced a new symbol, $\varphi$, for the phase, because we want to adopt 
the direction of circular rotation as the origin for $\gamma$; the relation between 
the two phases is  $\varphi = \phi + \pi/2$. In the figures, we represent the galactic
rotation in the clockwise direction, while $\gamma$ increases in the counterclockwise
direction. The situation $\gamma = 0^o$ occurs when the cluster is at its minimum galactic
radius,  along the epicycle trajectory ($\phi = -\pi/2$, in Equation 2).  

\begin{figure}
\plotone{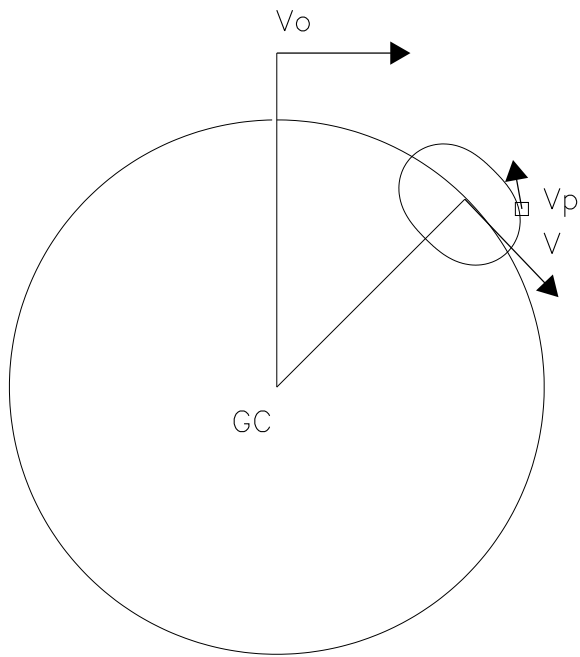}
\caption{Epicycle orbit around an equilibrium point rotating with circular velocity
V around the Galactic Center. V$_p$ represents the perturbed velocity.}
\label{figure1}
\end{figure}

\begin{figure}
\plotone{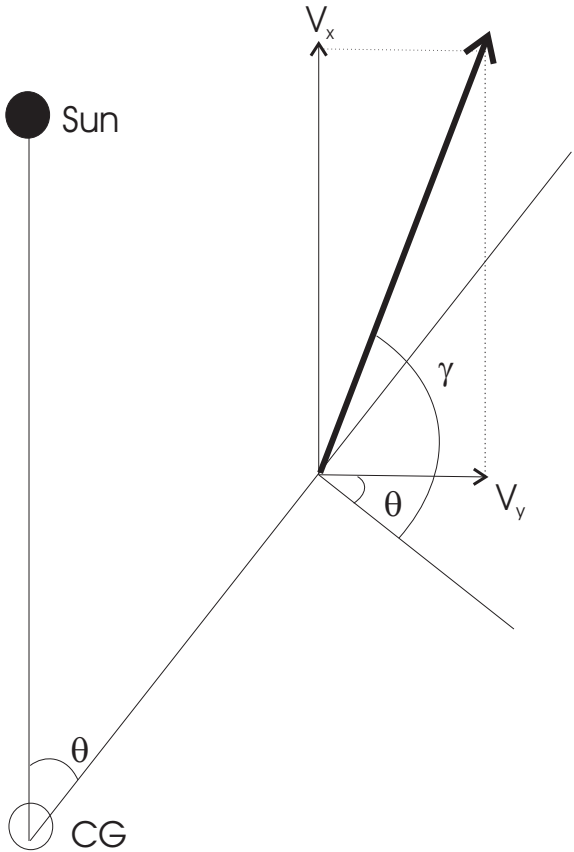}
\caption{The geometry of the velocity vectors. We adopt a usual convention with the Y axis 
oriented along the circular rotation at the solar position, and the X axis in the
galactic radial direction, oriented towards external regions. One can see that
$\gamma = \theta + tan^{-1} (V_x/V_y)$, with $\theta = sin^{-1}(y/R)$}
\label{figure2}
\end{figure}

In the hypothetic case $a=b$, $\gamma$  would increase linearly with time, with a slope
$\kappa$. In the case $a/b$ = 1.5, $\gamma$ is a smooth increasing function of time,
with a slope that on the average is equal to $\kappa$, since in a time $T= 2\pi/\kappa$,
$\gamma$ varies $360^o$. If all the clusters had a same starting angle, 
in a plot of $\gamma$ versus age of the clusters the slope of the fitted line would
give directly the value of $\kappa$. 

The present day velocity vector of an open cluster with respect to the Sun, projected  on
the galactic plane, is obtained from the data given in the catalog: it is the sum of the
radial velocity projected on the plane, and of the velocity in the plane  of the sky,
derived from  proper motion and distance,  also projected on the galactic plane. We
first convert the velocities to the Local Standard of Rest (LSR). Different sets of
solar velocity components ($u,v,w$) with respect to the LSR are recommended in the
literature, like the Standard Solar Motion, with ($u,v,w$) = (-10.4, 14.8, 7.3)
(in kms$^{-1}$, Mihalas \& Binney, 1981),  (-10.4, 5.3, 7.3) from the analysis of 
the Hipparcos data by Dehnen et al. (1987), (-9.7, 5.2, 6.7) also based on Hipparcos
data by Bienaym\'e (1999), or (-9.9, 12.1, 7.5) according to Abad et 
al.(2003), among others. Since differences of a few kms$^{-1}$ between the 
diverse systems only occur in the $v$ component, we varied it to verify how it
affects our results. We finally adopted  $v$ = 9.0 kms$^{-1}$, since this value 
corresponds to a minimum of the rms deviation of the adjusted points in the 
determination of the epicycle frequency. However, this choice is not critical for
our results; almost the same values of epicycle frequencies were 
obtained with $v$ in the range 6 to 12 kms$^{-1}$. Interestingly, we have 
another argument  based on the new ideas of the present paper in favor of this
intermediate value of $v$ between the extreme ones, 5 and 15 kms$^{-1}$. The
perturbation velocities are rotating, and since the clusters span a range of ages,
even if their initial velocities had some preferential direction, the present direction
must be almost random (or cover almost uniformly the range -180$^\circ$
to 180$^\circ$). However, if we adopt an artificially incorrect value for   $v$,
like 0 kms$^{-1}$ or 20 kms$^{-1}$, the present day distribution becomes
either very concentrated around $\pm$ 180$^\circ$ or around 0$^\circ$. This 
experiment tells us that the correct $v$ must be near the center of the range
 0 kms$^{-1}$- 20 kms$^{-1}$.

A further step in the treatment of the velocity vectors is their transformation to
the local reference frame of each cluster, as illustrated in Figure 2. The new frame 
has an axis in the local direction of rotation and the other in the local radial 
direction. This step includes the corrections for differential rotation, described 
below. The local frame of the clusters at present time is the most convenient one to 
start numerical integration of orbits. However, for the other use of the 
velocity vectors, which is the direct determination of the epicycle frequency, a
small correction is still required to place the velocity in the frame of reference 
of the guiding center, as later discussed. 

In this work, we do not take into account the velocities perpendicular to the 
plane; only the velocity components projected on the galactic plane are investigated.

\subsection{Corrections for differential rotation and local frame of reference}

After electing a rotation curve, we subtract from the observed velocity of a cluster
the velocity expected from pure circular rotation at its position; what is left
is the perturbation velocity. This is not different from what DL called correction
for differential rotation. The corrections can be made in different ways; the one that
we adopted is to compute, for the longitude and distance of each cluster, 
the expected  U and V velocity components, assuming that the cluster
presents pure circular rotation, and subtracting the  velocity components
(0, V$_0$) of the LSR. Of course, the perturbation velocity vectors that we 
derive (observed - expected velocities) depend slightly on the choice of the galactic parameters 
$R_0$, $V_0$, $(dV/dR)_0$ (the radius of the solar orbit, the rotation velocity of 
the LSR around the galactic center, and the slope of the rotation curve, respectively).
 However, we are dealing with second order (or differential) effects; if we adopt a
 rotation curve that is too low or too high, the LSR velocity and the expected
cluster velocity are affected in a similar way, and the computed perturbation velocity 
does not change too much. As already discussed, once a perturbation velocity vector 
is established, its angle $\gamma$ will be measured with respect to the local  
rotation direction.

Since the epicycle frequency depends strongly on the galactic radius (equation 1), we must
select the clusters in a narrow range of radius, to be able to observe a well defined
rotation of the perturbation velocity. We must make a compromise, since too narrow 
galactic radius ranges will contain only  a few clusters. After a number of experiments, 
we found that a radial extent of the order of 0.6 - 1 kpc is convenient. A special radius
range is the one that contains the Sun (let us say, the range  $R_0 \pm $ 0.4 kpc),
because it is not affected by differential rotation; whatever the rotation
curve we choose, the expected circular rotation velocity of the clusters is the same
of LSR rotation around the galactic center.

\subsubsection{Rotation curves}

For radius ranges (or rings) relatively distant from $R_0$, the choice of the rotation
curve is relevant.  Our task is risky, since we are willing to observe perturbations of
the order of 10 kms$^{-1}$ over a line-of-sight projection of rotation velocity 
of the order of 130 kms$^{-1}$.
For instance, if the true rotation velocity of the Galaxy in a ring is
larger than the one that we  use for the corrections, the perturbation velocities
are under-corrected in the direction of $\eta'$ and the observed $\gamma$s will tend
to be closer to zero than they should be. The effect is similar to the one that we
described previously, concerning the $v$ component of the solar velocity. In principle,
the fact that under-corrections or over-corrections tend to destroy the "normal" 
distribution of present day angles of the velocity vectors could be used as a guide
to select the correct rotation curve. But this is not the purpose of the present work. 

Anyway, in order to check the effect of the differential rotation corrections on
the derived epicycle frequency, we performed the calculations over a grid of rotation
curves. We varied $R_0$ from 7.0 to 8.5 kpc, in steps of  0.5 kpc, and $V_0$ from
170 to 250 kms$^{-1}$, in steps of 10 kms$^{-1}$. The method is very similar to that
of DL for the determination of the pattern rotation speed; for each point of the
grid we make use of a rotation curve that passes at the corresponding ($R_0$, $V_0$)
and  fits the observed CO data taken from Clemens (1985);  the same set of rotation
curves of DL was used here. These are curves directly derived from observations, 
with the rotation velocities recomputed for each adopted ($R_0$, $V_0$). A simple
and smooth analytical expression was fitted to each of these curves, in order to be 
able to compute the corresponding "theoretical" epicycle frequency. We refer to 
these curves as CO-based curves or smooth curves. We also performed a number of
experiments with constant velocity curves, or pure "flat"curves. 

Not all the points of this grid of rotation curves and radial ranges produce a valid 
measurement of the epicycle frequency. We later discuss in more detail an  example
with a specific choice of $R_0$, in order to clarify what we consider a valid
measurement. 

\subsubsection{The reference frame of the guiding center}

The equations 2 to 5 are valid in the reference frame of the guiding center of
the epicycle motion, which does not coincide with the local reference frame at present
epoch. A small correction to the angle $\gamma$, corresponding to a 
change of the velocity component in the $\eta$ direction, is required for this last
frame conversion. The velocity shift only affects the direction of circular rotation,
since  the local frame and the the frame of the guiding center have the same 
(zero) radial velocity.

The expression for the correction is derived in Appendix A, based on the equations
of motion  derived by  Makarov et al.(2004, hereafter MOT). It should be noted that the
 MOT's equations are mainly intended to describe the positions of stars as a
 function of time. In the analysis of positions, a secular drift appears and is
 quite important, this being the reason for the stretching of open clusters. In
the present work, we are not dealing with positions, but only with velocities 
(our analysis is based on plots of velocity directions as a function of age). Velocities
are not affected by secular terms like positions are; this is why young associations
that are considerably stretched (like Sco-Cen, and the stellar streams) are
easily recognized through their common spatial velocities. In our particular case
the present day position of the particles (clusters) coincide with the origin of
the individual local frame. As we move towards the past or  the future, 
the local frame rotates with a constant circular velocity and the particles move out
from the origin.

It is shown in Appendix A that the amplitude of the correction $\delta \gamma$ to
the present day angles, that we apply to each point on the graphs like figure 4, has 
an rms value about 15$^\circ$, and is on the average equal to zero. Since the tolerance
adopted in our analysis (section 3.3) for a point to be considered as belonging 
to a fitted line is 20$^\circ$, and the fitted parameters ($\kappa$ and 
$\varphi$) depend on many points, the errors that would be introduced by not
performing the $\delta \gamma$ correction would be  small. For such a small correction,
a first order estimation of it is largely sufficient. For this reason,
the fact that the correction is based on a theory that makes use of the
Oort's constants, of which we do not know a priory the exact values, is not relevant.
The use of different sets of Oort's constants produces minor differences in the
corrections (as illustrated in Appendix A), that do not affect the fitted parameters. 

\subsubsection{The modulus of the velocity perturbation}
The histogram of the modulus of the velocity vectors, or residual velocities after
correction for differential rotation,  is presented in Figure 3. The peak of the 
distribution is at about 10 kms$^{-1}$. In parts of the following study we remove
from the sample the clusters with modulus larger than 20 kms$^{-1}$. The  reason
is that if the amplitude of the perturbation is large, the simple epicycle
approximation is not valid. However, when we are dealing with numerical integration
of the orbits, which do not rely on the epicycle approximation, this restriction is 
not needed. We also remove from the sample the too small velocities 
($ \mid V \mid <$ 3 kms$^{-1}$ ), since the velocity components would be of the
same order of the measurement errors, and the error on the angle $\gamma$ could be large.

\begin{figure}
\plotone{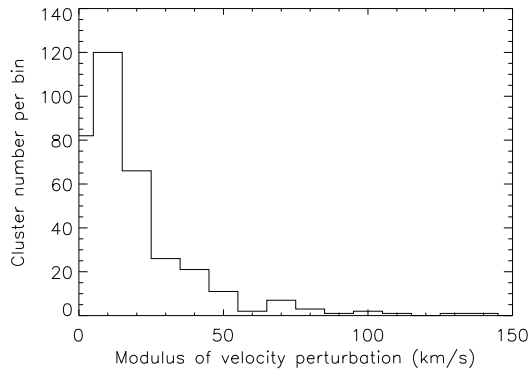}
\caption{Histogram of the modulus of the residual velocity vectors of the open clusters in
the galactic plane, after subtracting the normal rotation velocity at that radius.  
}
\label{figure3}
\end{figure}

\subsection{The case $R_0$=7.5 kpc, and clusters near galactic radius R$_0$}

The measurement of the epicycle frequency gives us a link between $R_0$ and $V_0$,
like the well known link between the same parameters given by the Oort's 
constants A and B. If we fix one of these parameters, we can derive the other. 
Recent work often adopts R$_0$ = 7.5 kpc (Racine \& Harris, 1989, Reid, 1993,
and many others). This shorter galactic scale, compared to the IAU recommended 
8.5 kpc scale, is supported by VLBI observations of H$_2$O masers associated 
with the Galactic center. The distance to the Galactic center derived from 
infrared photometry of bulge red clump stars (Nishiyama et al. 2006) is 
R$_0$=7.52 $\pm$0.10 kpc, while astrometric and spectroscopic observations 
of the star S2 orbiting the massive black hole in the Galactic center 
taken at the ESO VLT (Eisenhauer et al., 2005) gives 7.94 $\pm$ 0.42 kpc.

Figure 4 (a,b,c) are plots of the angle of the perturbed velocity vector
with the direction of circular rotation, as a function of age. In these
examples, the galactic positions of the clusters were computed considering that
R$_0$ = 7.5 kpc, and the plots a, b, c, show clusters selected in different
ranges of galactic radius. In particular, the range shown in Figure 4b is
 7.1 $< R <$ 7.9 kpc. As we already mentioned, this is a special radius range,
since the choice of the rotation curve has no influence on the  fitted $\kappa$.
The positions and velocities in the galactic plane of the same sample of clusters
are shown in Figure 5. In Figure 4a and 4c, the two lines   represent the
function (6), for a same value of $\kappa$, but  different values of $\varphi$. 
When a line reaches the maximum angle that can be measured  ($180^o$), 
 it starts again at $-180^o$. In Figure 4b only one line was plotted, as it seems 
 to be a dominant one, fitting alone 12 clusters, considering that a cluster 
 is fitted if it is situated within $\pm 20^o$, in the vertical direction, from
 the line. The points situated around this line are reproduced in Figure
 6 without the folding at multiples of 180$^\circ$; 2 points slightly off the 
 $\pm 20^o$ limit were included. The rms deviation of the points from the line
 is 14$^\circ$. It can be noted from Figure 4 that for ages larger than
 50 Myr, no points considered as not belonging to the line are closer than about
 130$^\circ$(or 9$\sigma$) from it. This justifies our method of "isolating" a 
 line to fit the theoretical function. In the present case the fitted value of $\kappa$
 is 42.0 $\pm $ 4 kms$^{-1}$kpc$^{-1} $; this is the epicycle frequency at the
 solar radius, one of the main results reported here. The initial angle is 
 61$^\circ$, in the frame of the guiding center, or a little more, about
 80$^\circ$ in the local frame of the clusters. This initial angle is in good
 agreement with what is obtained from a totally different method (Figure 4e). 
 The probability of the apparent alignment of points being a chance coincidence
 is discussed in the next section. 

\begin{figure*}[t]
\plotone{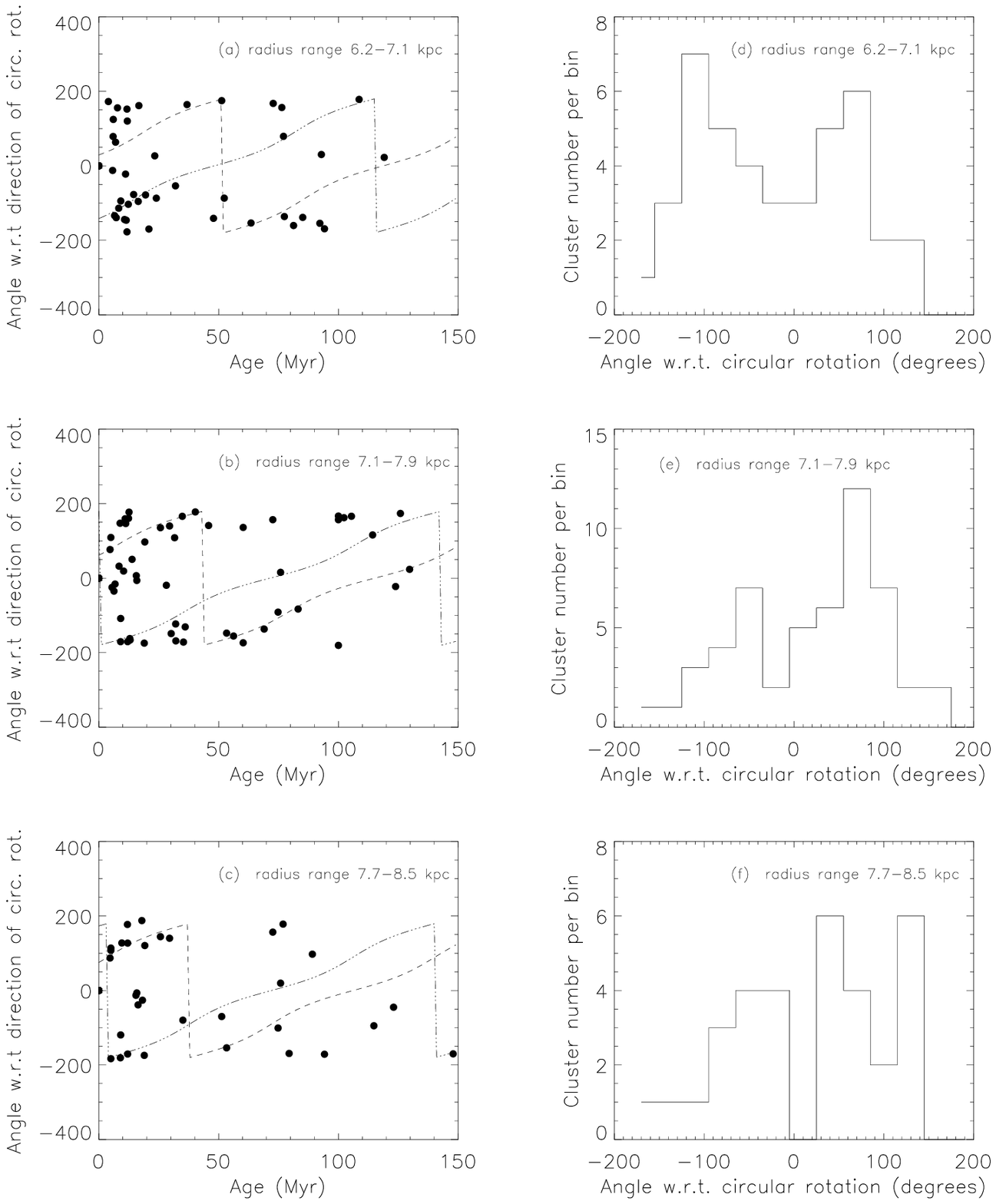}
\caption{Left-side (a,b,c): angle $\gamma$ between the perturbed velocity and the
 direction of circular rotation as a function of the age of the clusters. The 
 lines represent the function (6) for different values of $\varphi$. The clusters
 were selected in different ranges of galactic radius, indicated in each frame. 
 The fitted values of $\kappa$ were 46, 42 and 45 kms$^{-1}$kpc$^{-1}$ 
 respectively. The right side frames are the histograms of initial velocities for
 the corresponding radius ranges, obtained by numerical integration of the orbits
 (section 3.4)} 
\label{figure4}
\end{figure*}

\begin{figure}
\plotone{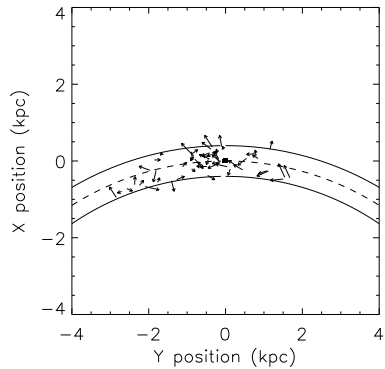}
\caption{Observed position and perturbation velocity vectors of the clusters in 
the range of galactic radius 7.1$< R <$ 7.9 kpc. The  radius range is indicated
by full lines and the solar radius by a dotted line. In the text we refer to this
region as the R = 7.5 kpc ring. The galactic center is not contained in the figure.}
\label{figure5}
\end{figure}

\begin{figure}
\plotone{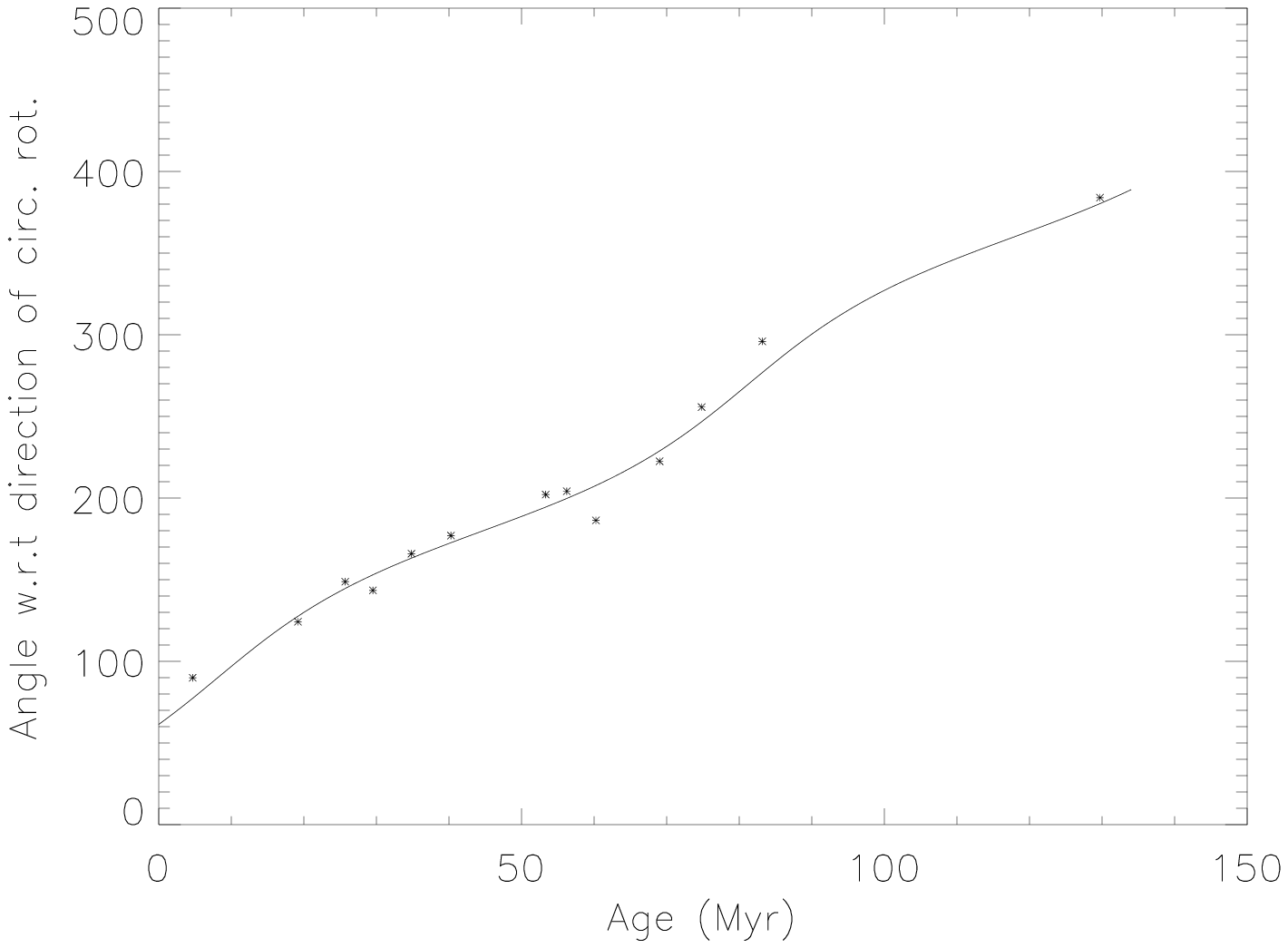}
\caption{Angle $\gamma$ between the perturbed velocity and the direction of circular
rotation as a function of the age of the clusters, for the clusters situated around
the line shown in Figure 4b, with the 180$^\circ$ folding removed. The adjusted line
gives $\kappa$ = 42.0 $\pm $ 4 kms$^{-1}$kpc$^{-1} $ and $\varphi$ = 70$^\circ$, which 
corresponds to an initial angle 61$^\circ$ (Equation 6).}
\label{figure6}
\end{figure}

The fact that a reasonable number of points fall on a same line (for instance, 
12 clusters within $\pm$ 20$^o$ of the line that begins at 61$^o$) reveals that some
process confines the direction of the initial velocity of the clusters to a same 
angle with respect to the direction of circular rotation, for periods of time of 
more than 100 Myr. We postpone for the moment the discussion on the nature of such
process.

\subsection{Probabilities of alignment in a random process}

In the case illustrated in Figures 4b, the sample contains 52 clusters.
The probability of 12 clusters being fitted by a theoretical line, as it happens 
in this case, is small. The probability of a point falling on a given line which
has a "thickness" of 40$^o$ is 40/360 = 1/9. The probability of 12 points being 
on a given line, out of 52 experiments, or 12 successes and 40 failures, is 
[52!/(12! 40!)](1/9)$^{12}$(8/9)$^{40}$, $\approx$  7$\times 10^{-3}$.

In the example of Figure 4a, with a sample of 40 points, 2 lines have been fitted,
with a total of 16 points on them. In a random process, the probability of
falling on one of the two lines is 2/9. If  40 points are randomly placed on
the diagram, the probability of 16 points  falling on any one of 2 given lines
(15 successes and 21 failures) is 
 [40!/ (16! 24!)](2/9)$^{16}$(7/9)$^{24}$ $\approx 5\times 10^{-3}$.
 
 In the first example, one could argue that the problem cannot be reduced to that
 of throwing 52 points on graph and computing the probability of 12 of them being within
 a given distance of a line, since the line was not previously present;
 we fitted their parameters so that they  pass through a group of points that
 resulted to be aligned. This is equivalent to reduce the number of "successes", 
 for instance, we can choose 2 points to define the phase and slope of the  line.
 In this case we require only 12  successes and 40 failures; the probability
 is still only about 3$\%$. In reality, the probability  of the observed 
 alignments of points being spurious can be discarded, since the fitted
 value of $\kappa$ are always very close to the ones expected from theory,
 and the initial phases are confirmed by another method. Furthermore, there is 
 a continuous variation of $\kappa$ from one ring to the other; for instance,
 the clusters in  figures 4a and 4b are not the same, since the galactic rings do
 not overlap, but the fitted slope changes smoothly from one ring to the other. 
 We note that the choice of the number of fitted lines  is arbitrary; similar
 conclusions can be obtained with 1, 2, or 3  lines. We also verified that
 our conclusions do not depend strongly on the  "thickness" of the lines; if we
 decide that points at distances up to 15$^o$ or 25$^o$ from a line are fitted
 points, the number of fitted points varies, but the change in the probability 
 of being on a line by chance tends to compensate the variation, in the calculation
 of probability.
 
 The simple analysis of probabilities tells us that we are not dealing with random 
 alignments, but also warns us that we should not attempt to draw conclusions
 from cases with a small number of aligned points, like for instance less
 than half of the total number of points being on 3 candidate lines, or less than
 one third of the points being on two lines, within a tolerance of 20$^o$. 
 When the number of clusters decrease, or when the galactic parameters
 or rotation curves are different from the correct ones, we are
 sometimes confronted with situations in which  very different values of $\kappa$ 
 seem to fit the points with about the same quality of the fit. This limits the
 number of galactic rings and the range of parameters  that can be explored. 
 
 \subsection{Numerical integration of the orbits and histograms of initial directions}
 
The approximation described in the previous sections is  useful to provide
an estimate of the epicycle frequency without depending on a precise 
knowledge of the rotation curve. A rough idea of the distribution of initial 
velocity directions is obtained simply by looking at the starting angles of
the fitted lines (at age=0). 

To obtain the distribution of initial angles, a second method is 
to perform a numerical integration of the stellar orbits. For this integration,
a first step is to obtain the axis-symmetric potential of the Galaxy, by integrating the
force $V(R)^2/R$ based on the adopted rotation curve. In this way, the potential
does not depend on any assumption on the different components of the Galaxy
(disk, bulge, etc.). The present day position and velocity of a cluster define a 
quadrivector q(t) containing the polar coordinates, the radial velocity
and the angular momentum, which are the initial conditions of the integration. 
In the integration, at each time step, $q(t+dt)$ is deduced from
$q(t)$ using a Runge-Kutta IV method. The value of $dt$ is chosen so as to 
have about  360 points along a circle around the Galactic center. The integration
is performed backward in time, for a total time equal to the age.

If the adopted rotation curve is close to the true one, the two methods
should produce similar initial velocity distributions. Therefore, the comparison 
between the results of the two methods is a  tool to make a choice among
different rotation curves. As  examples of  histogram of initial velocity 
directions, those obtained with the same samples used in Figure 4 (a,b,c), are shown in 
Figure 4 (d,e,f, respectively).  Note that we expect to obtain similar, but not
precisely the same distribution of initial velocities with the two methods, since 
the methods rely on different simplifying assumptions. For instance, the numerical
integration takes into account the present day galactic radius of individual clusters,
while the slope-fitting method considers intervals of galactic radius, and supposes that 
the equilibrium radius of the circular orbit is approximately the center of 
the radius range. It should be remembered that the initial angle in the slope-fitting
method is in the frame of reference of the guiding centers (Section 3.1.2). 

The agreement between the two methods is found to be quite satisfactory when rotation
curves with high values of $V_0$ (220, 230 kms$^{-1}$kpc$^{-1} $) are used, but become
poor with small $V_0$, like 180 kms$^{-1}$kpc$^{-1} $.
 
 \subsection{Possible sources of errors}
 
 A source of uncertainty in the value of $\kappa$ that we obtain, in the examples
 that we discussed,  illustrated in Figure 4, is that $\kappa$  depends on the 
 definition of the LSR. It may happen that we are not using the best set of solar
 velocity components for the LSR corrections, in the choice described in section 3.
 The "correct" LSR would be the reference frame that rotates around the
 Galactic center with the velocity of the rotation curve at R$_0$. It can be seen
 from Figure 2 that if we add a correction to $V_y$ (in the direction of rotation),
 the computed value of $\gamma$ will change. A small change in the LSR
 definition produces a same $\delta V_y$ for all the clusters, but different angles
 $\gamma$ are affected in a different way, so that the slope of the fitted 
 lines in Figure 4 may change. The arguments which led us to adopt $v$ = 9 kms$^{-1}$
  are explained in section 3.0. We verified that typically, if we add 1 kms$^{-1}$ 
  to the $V_y$ component of each cluster, to mimic the effect of an error on
  the $v$ component of the solar motion,  the fitted $\gamma$ is changed about
  1 kms$^{-1}$kpc$^{-1}$. The error on $\kappa$ introduced by the uncertainty in $v$
  is about $\pm$3 kms$^{-1}$kpc$^{-1}$.
 
 It was mentioned in Section 3 that the ratio a/b is approximately 1.5, but the precise 
 value depends on the adopted $\Omega$ and $\kappa$. A number of tests showed that
 a small variation of this ratio (like for instance taking a/b = 1.4) does not produce
 any change in the fitted values of $\kappa$, so that it was not even necessary to use 
 an iterative procedure to determine $\kappa$. Similarly, in section 3.1.2 we commented
 that the correction $\delta \gamma$ for the frame of reference of the guiding centers
 depend on the Oort's constants, which in turn depends on the adopted rotation curve.
 In this case too, these are second order effects, and an iterative procedure is not 
 required. In principle, when a line is fitted to the observed points like in Figure 6,
 the scattering of the points due to the errors in radial velocities and proper motions
 contained in the catalog, as well as the small effects that we just mentioned,
 are taken into account in the error on the slope given by the least-square fitting.
 We consider that our final errors on $\kappa$ are $\pm$4 kms$^{-1}$kpc$^{-1}$.
 
 Other effects like  the possible  perturbation of the orbits by the spiral structure
 (see eg L\'epine et al, 2003), or the effects of the local wiggles of the rotation
 curve (section 3.8) are not considered here as errors of measurements, but the cause
 of discrepancies between the measured  $\kappa$ and theoretical values expected
 from too smooth rotation curves.
 
 \subsection{Detailed analysis and the derived value of $V_0$}

We expect that if a rotation curve is a good approximation of the true one,
it will produce a match between the theoretical and observed $\kappa$, not
only near $R_0$, but at other galactic radii as well. In Figure 7 we present 
$\kappa$ as a function of radius, obtained from the same slope-fitting method
illustrated in Figures 4(a,b,c). We remark that all the determinations of 
$\kappa$ rely on this method, the numerical integration is used only in the 
statistics of initial velocities. The method does not give reliable results between
8.0 and 8.9 kpc, because the number of clusters becomes too small, and in addition, the 
number of initial phases seem to be larger. Only at 8.9 kpc we found again a
reliable result. Although the observed epicycle frequencies do not depart much 
from those predicted (based on Eq. 1) by a smooth rotation curve with $V_0$=
220 kms$^{-1}$, the slope of $\kappa$ as a function of radius is steeper than that
of  smooth rotation curves. The data in Figure 7 was also fitted by a smooth $V_0$ =
230 kms$^{-1}$ rotation curve modified by the addition of a Gaussian  minimum
centered at 7.8 kpc and peak value 6 kms$^{-1}$ , the corresponding value of $V_0$ 
being about 225 kms$^{-1}$. The  existence of a minimum is further discussed in a
 next section. We do not intend to present a precise fitting of it, but only to
 illustrate that it could be an explanation for the anomalous slope of  $\kappa$.
 Such a minimum is almost not perceptible in the rotation curve (Figure 10).

\begin{figure}
\plotone{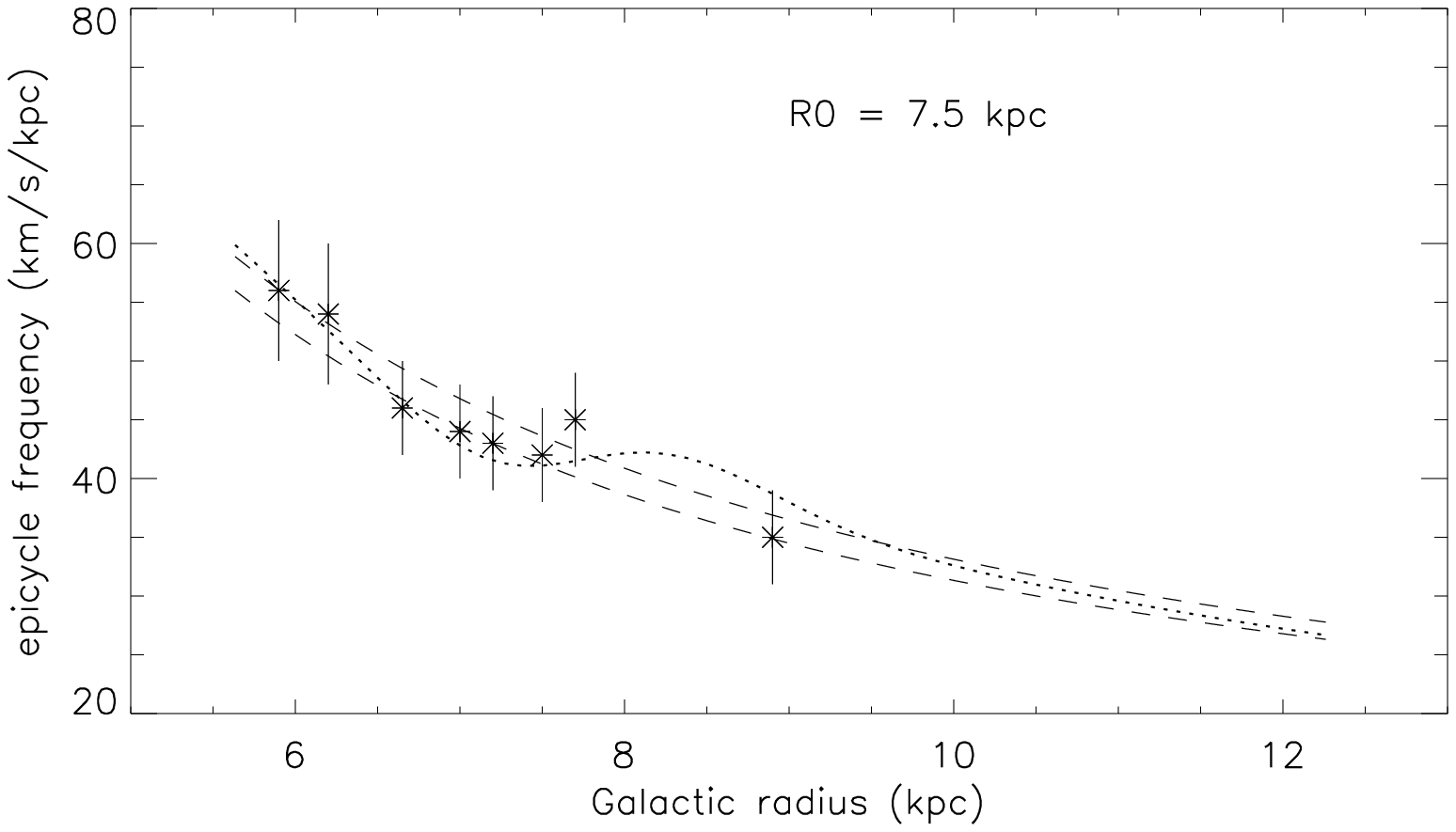}
\caption{Epicycle frequency measured at different galactic radii. The dashed
lines represent the theoretical $\kappa$ corresponding to smooth rotation 
curves with $V_0$ =  230 kms$^{-1}$(upper curve) and 220 kms$^{-1}$ (lower curve)
both with $R_0$ = 7.5 kpc. The dotted line correspond to the theoretical $\kappa$ 
obtained with a smooth rotation curve with $V_0$ = 232 kms$^{-1}$, to which 
a Gaussian minimum centered at 7.8 kpc has been added.} 
\label{figure7}
\end{figure}  

 Interesting results are obtained by varying the rotation curves with a fixed $R_0$.
 Figure 8 presents the theoretical and observed epicycle frequency as a function
 of the adopted $V_0$, for a fixed $R_0$= 7.5 kpc. What we call here the 
 theoretical $\kappa$ is the value obtained from direct use of equation (1),
 with $\Omega(R)$ given by the adopted rotation curve; the observed $\kappa$
 is obtained from the slopes of lines in plots similar to those shown in Figure 4.
 For each $V_0$, the same rotation curve is used to determine the theoretical
 frequency and for the differential rotation corrections. Figure 8 turns clear
 how the best value of $V_0$ can be derived, and how the uncertainty on $\kappa$
 affects this $V_0$ value.
 
 Let us first analyse the case of pure flat rotation curves (V($R$)=V($R_0$)= 
 constant). For such curves, it can be derived from equation (1) that  $\kappa$ = 
 $\sqrt2 \Omega$. At $R_0$, $\Omega$ = $V_0 /R_0$; the dashed line in Figure 8
 gives $\kappa$  as a function of $V_0$, for flat curves with R$_0$ = 7.5 kpc.
 This line cuts the line of observed $\kappa$ at 42 kms$^{-1}$kpc$^{-1}$ 
 and V$_0$ = 226 kms$^{-1}$. In other words, if the rotation curve were a flat
 one, our measurement of $\kappa$ = 42 $\pm$ 4 kms$^{-1}$kpc$^{-1}$ would imply
 that V$_0$ is 226 $\pm$ 15 kms$^{-1}$.

 \begin{figure}
\plotone{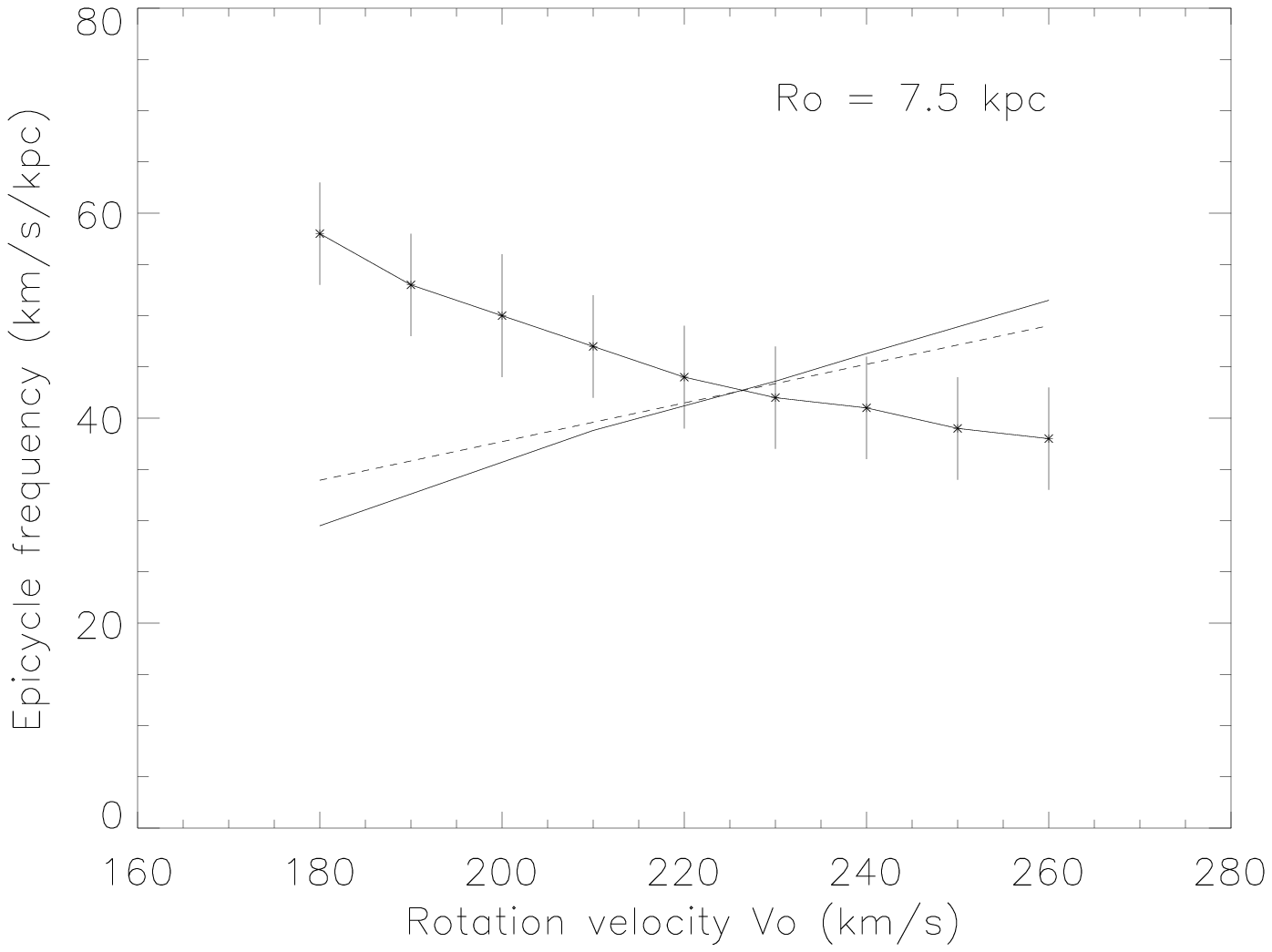}
\caption{The epicycle frequency as a function of $V_0$ for different rotation
curves, all based on $R_0$ = 7.5 kpc. The full line with positive slope
is the theoretical prediction for the series of CO-based rotation curves  
with different values  of $V_0$ (see section 3.1.1). The dashed line
is the theoretical  $\kappa$ for pure flat curves with different $V_0$. The
full line with negative slope shows the observed values of $\kappa$, based on 
angle versus age fitting, like in Figures 4(a,b,c). The theoretical and observed
$\kappa$  intersect at at about 226 kms$^{-1}$.}
\label{figure8}
\end{figure}

Besides the pure flat (slope zero) curves, another family of curves that we 
consider is that of smooth curves, obtained by fitting the CO data 
with different values of $V_0$ (section 3.1.1). The line representing the
theoretical $\kappa$ for this family of curves  also cuts the observed 
line at 226 kms$^{-1}$. The coincidence of the two "theoretical" lines
intersecting at 226 kms$^{-1}$ is not surprising, since the smooth curves 
for $V_0$ about 220-230  kms$^{-1}$ are  very close to flat ones
(see Figure 10).

\subsection{Rotation curves with $R_0$ = 7.5 kpc at different galactic radii}

We  present in Figure 9  the observed values of $\kappa$
(based on figures similar to Figure 4) and theoretical values of $\kappa$
(derived from the family of CO-based curves) for two distinct ranges of radius,
one smaller than $R_0$, 5.9 $< R < 6.5$ kpc (dashed lines), and the other 
larger than $R_0$, 8.2 $< R < 9.8$ kpc (dotted-dashed lines). The lines with positive
slope are the theoretical ones. The two lines corresponding to $R$ = 6.2 kpc 
intersect at about $V_0$ = 220 kms$^{-1}$; while the two lines corresponding to 
$R$ = 9.0 kpc intersect at about 235 kms$^{-1}$; both values have errors $\pm$ 10 kms$^{-1}$
and are in agreement   with $V_0$ = 226 kms$^{-1}$, previously obtained from the 
clusters situated near $R$ = 7.5 kpc, in Figure 6. 

\begin{figure}
\plotone{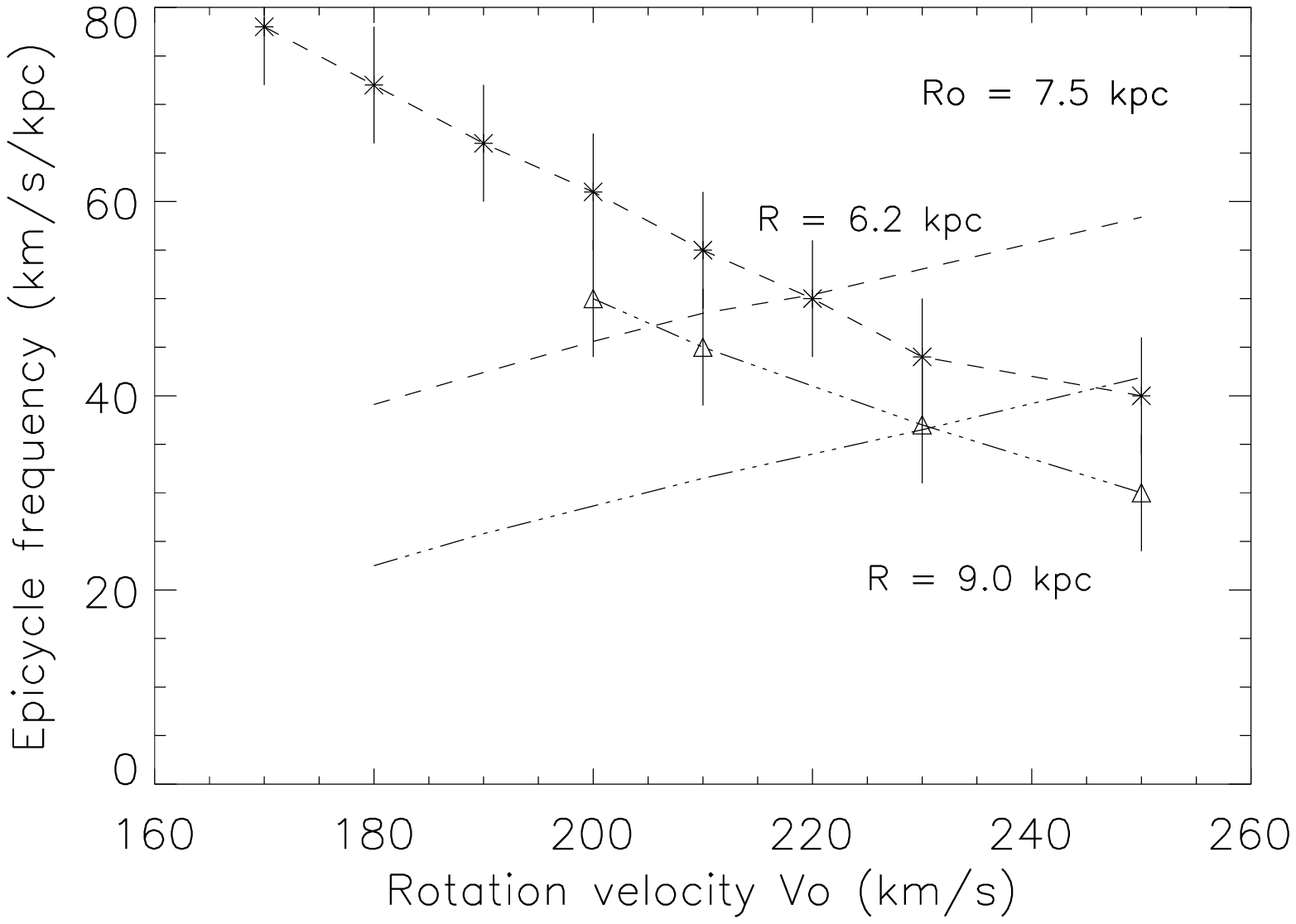}
\caption{Theoretical and observed values of $\kappa$ as a function of $V_0$
of the CO-based rotation curves (all with $R_0$ = 7.5 kpc), for two distinct
galactic radii: 6.2 kpc (dashed lines) and 9.0 kpc (dashed-dotted lines). The 
positive slope lines are the theoretical ones, and the negative slope are 
the observed ones, like in Figure 8.}
\label{figure9}
\end{figure} 

\subsection{Rotation curves with $R_0$= 7.0 , 8.0 and 8.5 kpc }
 
 An analysis similar to that of the previous subsections has been performed, 
 adopting $R_0$ = 7.0 kpc, $R_0$= 8.0 kpc and $R_0$= 8.5 kpc. For each adopted
 $R_0$, families of smooth rotation curves  with different values of $V_0$
 (as explained in section 3.1.1) were used to derive the theoretical $\kappa$
 at the radius $R_0$. On the other hand, the angle of the observed velocity
 perturbation with respect to the direction of circular rotation have been 
 plotted,  using the samples of clusters in the radius range $R_0~ \pm$ 0.4 kpc.
 The study was also made for rings that do not coincide with $R_0$; the width
 of the rings was always about $\pm$ 0.4 kpc. In all the cases the figures 
 look similar to Figure 9, with a rising theoretical line and a decreasing
 line of "observed"  $\kappa$. The intersection of the two lines is considered
 as the best estimate of $V_0$ for the given $R_0$. 
  The results are summarized in Table 1.

\begin{table}[t1]
\centering \caption[]{\label{tab1}Values of $V_0$ for which the observed 
and theoretical $\kappa$ are equal, for different adopted $R_0$ and different
radii of the galactic ring.}
\begin{flushleft}
\begin{tabular}{ccccccccc}
\hline $R_0$& ring radius & best $V_0$ & $\kappa$   \\
  kpc    & kpc     &  kms$^{-1}$ & kms$^{-1}$kpc${-1}$ \\
\hline \noalign{\smallskip}
7.0 & 5.7 & 195   & 50 \\
7.0 & 7.0  & 230 & 47   \\
\hline \noalign{\smallskip}
7.5 & 6.2   & 220  & 48  \\
7.5 & 7.5   & 225  & 42  \\
7.5 & 9.0   & 230  & 35  \\
\hline \noalign{\smallskip}
8.0 & 6.7   & 225  & 48 \\
8.0 & 8.0   & 230  & 40  \\
\hline \noalign{\smallskip}
8.5 & 7.3   & 210   & 42 \\
8.5 & 8.5   & 245   & 42 \\
\noalign{\smallskip} \hline
\end{tabular}
\end{flushleft}
\end{table}

 It can be seen that the observed $\kappa$ at $R$(ring)= $R_0$ is almost the same for
 the different adopted $R_0$ (the average of the $R$= $R_0$ rings in Table 1
 is $\kappa$=43 ). This is an expected result since the samples of 
 clusters are  the same in these cases, as the positions of the clusters are given
 relative  to the Sun. And, as we already mentioned, when we study a ring at $R_0$,
 the rotation curve has only a second order effect in the determination of $\kappa$.
 The fact that a same  best estimate of $V_0$ is obtained for $R_0$= 7.0, 7.5 and 8.0
 kpc is an indication that this value is a robust result.
 Furthermore, the fact that for $R_0$ = 7.5 and 8.0 kpc almost the same value of $V_0$ is
 obtained from two different rings is an indication that the smooth rotation
 curves that we are using in these cases are close to the real ones. 
 
 \subsection{A wiggled rotation curve and the Oort's constants}
 
  It is well accepted that the rotation curve presents irregularities in the 
  solar neighborhood. According to Olling and Merrifield (1998), there is a minimum 
  at about $R_0$ + 3 kpc, and a step to higher rotation velocities at about $R_0$ + 4 kpc. 
  Honma and Sofue 1997) used the same method of Merrifield (1992), and both groups obtain a
 step-like increase of rotation velocity, but at different radius: about 1.2 $R_0$ 
 (HS) or about 1.5 $R_0$ (M). Brand \& Blitz (1993) found the deepest minimum at
 about 1.15 $R_0$, and the start of a step-like increase at about 1.3 $R_0$. Amaral
 et al. (1996), based on an analysis of OH/IR stars, found the minimum at 1.07 $R_0$.
 We propose the curve presented in Figure 10, which contains a minor wiggle
 represented by a Gaussian minimum centered at 7.8 kpc with a half-width of 1kpc,
 and depth 8 kms$_{-1}$. Such a wiggle produces a reasonable fit of the epicycle
 frequency (see Figure 7), and the general curve  fits the Clemens (1985) data,
 excepting for the minimum at about 8.5 kpc. It should be noted that Clemens did 
 not use the same VLSR transformation that we adopted in the present work, and that 
 the we applied corrections for the different $R_0$ that are only strictly valid 
 for the data inside the solar circle.
 
 In most of the descriptions that we mentioned, the minimum is at 1.1 $R_0$ or beyond.
 So up to  $R_0$ the curve is still quite flat, and we believe that our analysis of
 the sample of clusters near $R_0$ is not seriously affected by the wiggles.
 Indeed, when we plot the angle $\gamma$ as a function of age, like in Figure 4, but
 using our  curve with aminor wiggle to perform the differential rotation correction, 
 the result looks very similar to Figure 4, except that the best fit of the slope of the lines
 gives us 41 kms$^{-1}$kpc$^{-1}$, instead of 42 kms$^{-1}$kpc$^{-1}$. This difference 
 is well within the error bars.

\begin{figure}
\plotone{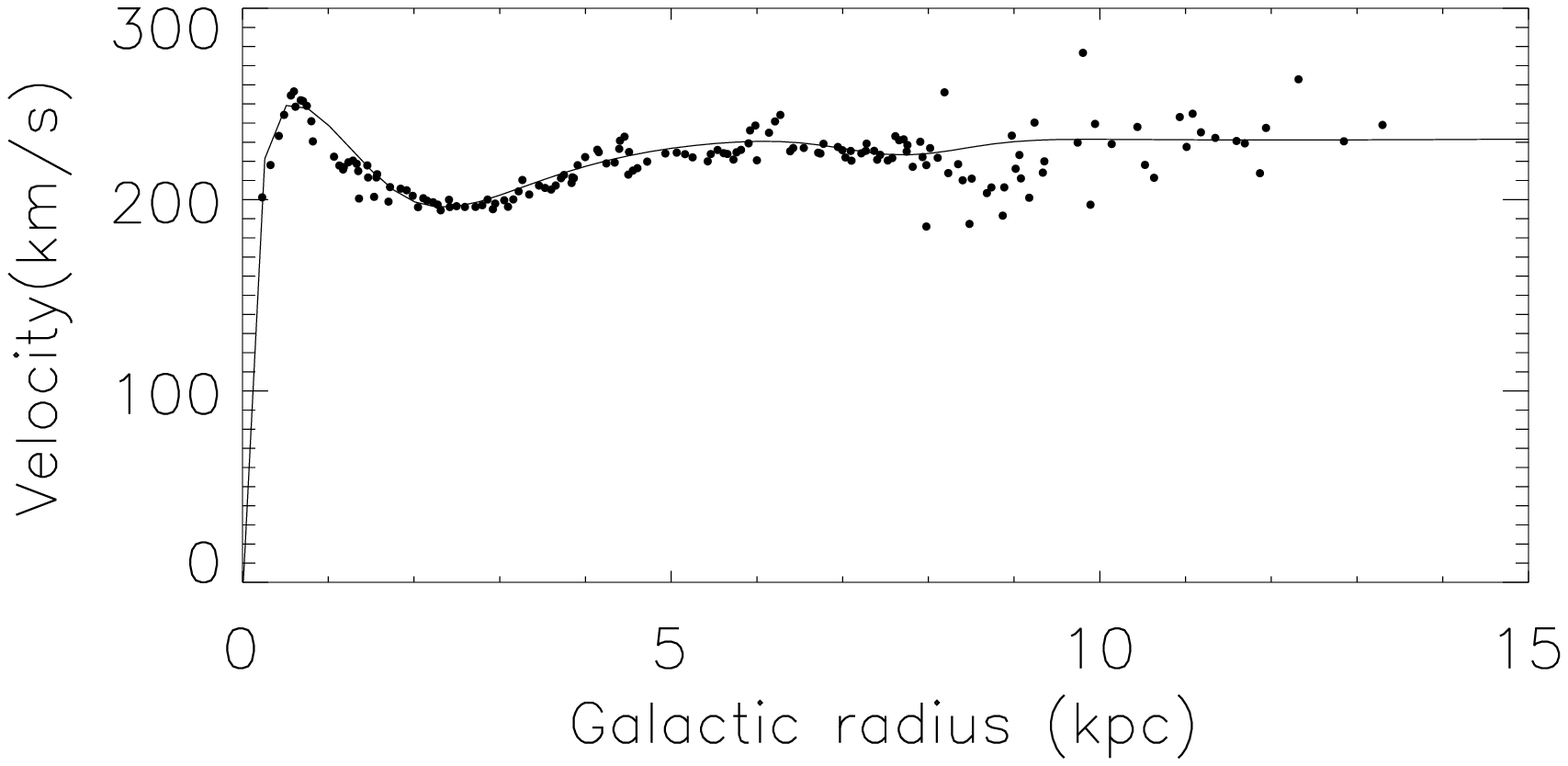}
\caption{The rotation curve of the Galaxy for $R_0$ = 7.5 kpc. 
The line is a fit to the CO data, using the empirical expression 
V= 240 exp(-(3.6/r)$^2$) + 350 exp(-r/3.5 -0.12/r); the units are
 km$^{-1}$ and kpc. A very small wiggle was added, represented by
a  Gaussian minimum -8 exp(-[(r-7.8)/1.0]$^2$), in order to better
fit the epicycle frequecy data (Figure 7). Including this term,
$V_0$ = 226 kms$^{-1}$ precisely at $R_0$, but outside the minimum
$V$ =230  kms$^{-1}$.}
\label{figure10}
\end{figure}

An important aspect of the minimum in the curve is that it could explain the apparent
contradiction between our results and previous results based on the Oort's constants.
The value of $V_0$ that we obtain may seem too high, since it corresponds to 
$V_0 /R_0$ = 30.1 (the units are  kms$^{-1}$kpc$^{-1}$ in what follows). In a first
 order approximation, the parameters V$_0$ and R$_0$ are linked together through 
 the Oort's constants A and B, which are determined by observations (V$_0$/R$_0$ 
  = A - B,  (dV/dR)$_{R0}$ = -A-B ). A value of A-B  which seemed  to be well 
 established is 26, confirmed by recent observations of a different nature 
 (Kalirai et al., 2004) which give 25.3. Feast \& Whitelock (1997) obtained 27.2 based
 on Hipparcos proper motions of Cepheids. However, Olling and Merrifield (1998)
 verified that the Oort's constants A and B differ significantly from the general
 $V_0 /R_0$ dependence, in the solar neighborhood. They attributed this effect to
 an anomaly in the local gas distribution. Olling \& Dehnen (2004) showed that 
 the most reliable tracers of the "true" Oort's constants are the red giants, and derived
 A-B = 33. Another previous determination with high value of A-B is that of 
 Backer \& Sramek (1999), who measured the proper motion of Sgr A, at the Galactic
 center, equal to 6.18 $\pm$ 0.19 mas yr$^{-1}$, equivalent to 29.2 $\pm$ 0.9. This
 is an interesting result since it does not depend on local wiggles. Branham (2002) 
 obtains $\Omega$= 30.3,  Fern\'andez et al. (2001),$\Omega$ =30, and Miyamoto and
 Zu (1998), $\Omega$ = 31.5,  all of them  from the kinematics of OB stars. Meztger et
 al. (1998) obtained 31  from Cepheid kinematics, and Mendez
 et al. (1999) obtained 31.7 from the Southern Proper Motion Program. We conclude 
 from this discussion that the value of $V_0$ that we obtain, and our choice
 $R_0$ = 7.5 kpc, are not in contradiction with the constraints of the Oort's constants.
 
  As we already mentioned, one of the main properties of the wiggles is a minimum 
  at about 1.1-1.2 $R_0$, observed by different authors and using different tracers.
  We only insist here on the existence of the minimum because it has not always been
  recognized, possibly because there seems to be no obvious reason for it. 
 The existence of this sharp (half-width about 1.0 kpc) minimum could  possibly
 be related to corotation. This resonance is a frontier between two
 regions of the galactic disk, one in which the matter flows towards the center, and
 one in which the matter flows towards external regions. This tends to form a "vacuum" 
 around corotation, as confirmed by hydrodynamic simulations (eg. L\'epine et al. 2001). 
 If less gas is present, less stars form, and after a few billion year a "gap" ring
 with a smaller stellar density forms. 
 
\subsection{Statistics of initial directions}

Typically, a ring with a width about 0.8 kpc contains 3 or 4 main directions 
of initial velocity perturbation, seen as peaks in the histograms. A preliminary
analysis of the position on the galactic plane where open clusters were born,
indicates that at least two different directions can originate from a same
position in the galactic plane, at a same epoch. It was shown by DL that the
clusters were born in spiral arms. We can now say that a same spiral arm can produce star
clusters going in different directions. However, this kind of analysis is difficult
to perform, since the samples become very small when narrow bins of ages and initial
positions are considered.

Different rings may have different main directions, but some general tendencies 
can be observed. Figures 11 to 13 present the distribution of initial velocities
in the local (U,V) frame of each cluster, for different ranges of radius, obtained
by numerical integration. One is the local region, from 7.1 to 7.9 kpc (Figure 11),
which has already been studied in detail, and is supposed to include the corotation
radius. Another region, from 5.0 to 7.1 kpc (Figure 12) is inside the  corotation
circle. We consider a large galactic radius interval to obtain better statistics.
 Finally, in Figure 13 we show the distribution for clusters in the radius range 
8.1 to 12 kpc, external to the corotation circle. One can see some preferential 
or non-random directions, which are the basis of the present work, since they 
offer the possibility of measuring the epicycle frequencies. Although the statistics
are poor, it does not seem that a description in terms of pure ellipsoids
would be an adequate one.

 Clear differences appear between the three regions, with dominant positive U velocities at
large galactic radii, and dominant negative U velocities in the internal region.
There seems to be a change of about 180$^\circ$ in the main velocity direction in the
comparison of internal and external regions (from -50$^\circ$ to + 130$^\circ$). 
This is in general agreement with the idea that outside corotation, the spiral arms 
tend to pull the galactic material towards the exterior, while inside corotation 
the arms pull the material towards the center.

In principle, we might expect that the clusters start their life traveling along the
spiral arms, so that they would not leave the arms before the brightest stars
reach the end of their life. This would contribute to the well defined visible
aspect of the arms. DL showed that if we select clusters with ages smaller than
20 Myr, the spiral structure appears clearly, while for ages larger than 30 Myr, 
it almost disappears. This time scale is comparable to the time required for a 
velocity vector to change its direction by 90$^\circ$, which is 1/4 of the epicycle
period. Interestingly, the dominant initial directions for the internal regions
of the Galaxy satisfy the condition above. If we consider the vector sum of the 
circular velocity and of the perturbation velocity, for instance a vector with modulus
about 230 kms$^{-1}$ in the V direction plus a perturbation with modulus 15 kms$^{-1}$
in the direction -70$^o$ or in the direction -140$^o$ (main directions seen in
Figure 12), the resulting velocity is approximately along the spiral arms,
considering a typical pitch angle of about 14$^o$. 

\begin{figure}
\plotone{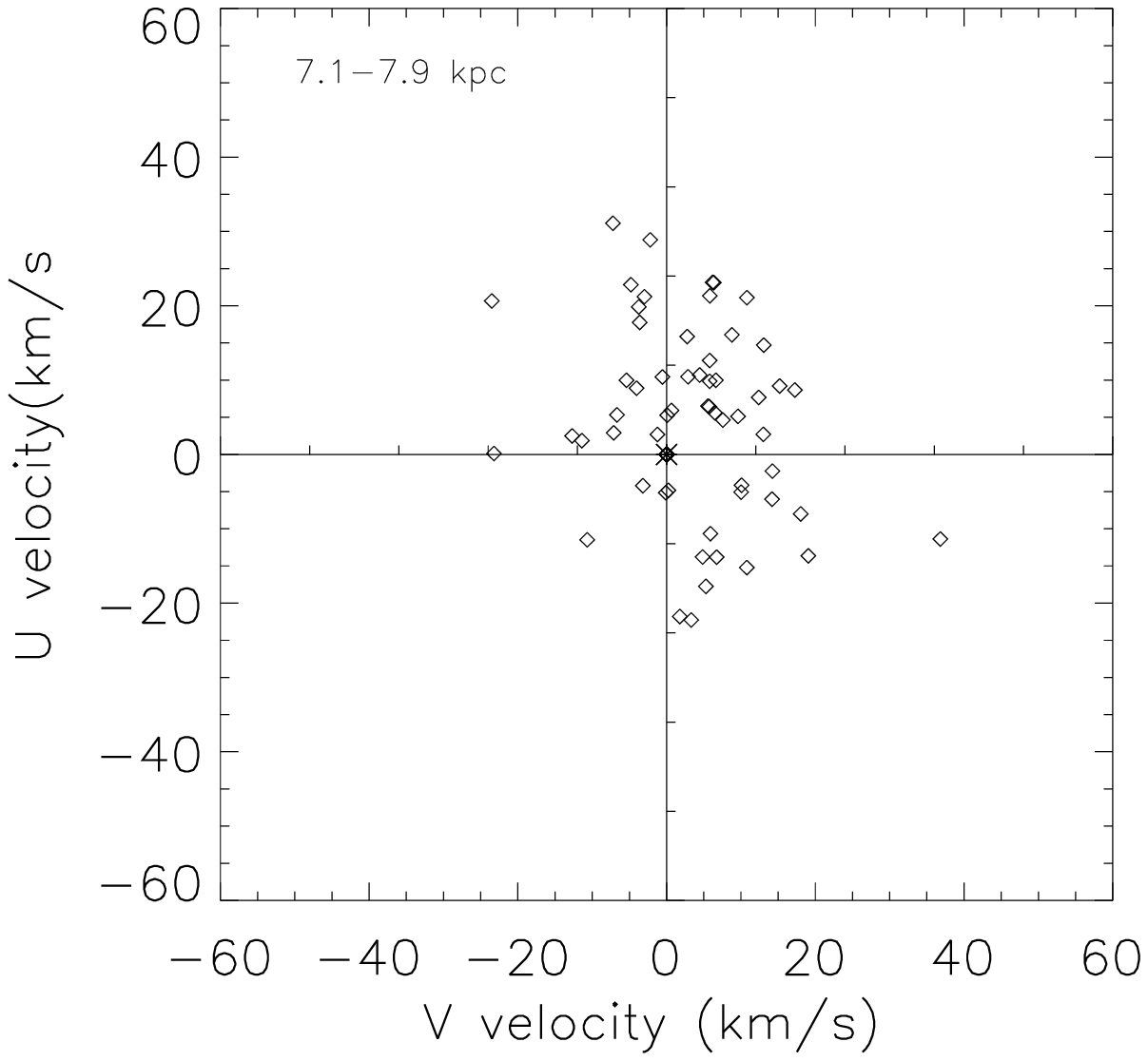}
\caption{Distribution of initial velocities of clusters in the galactic radius range 7.1 to
7.9 kpc}
\label{figure11}
\end{figure}

\begin{figure}
\plotone{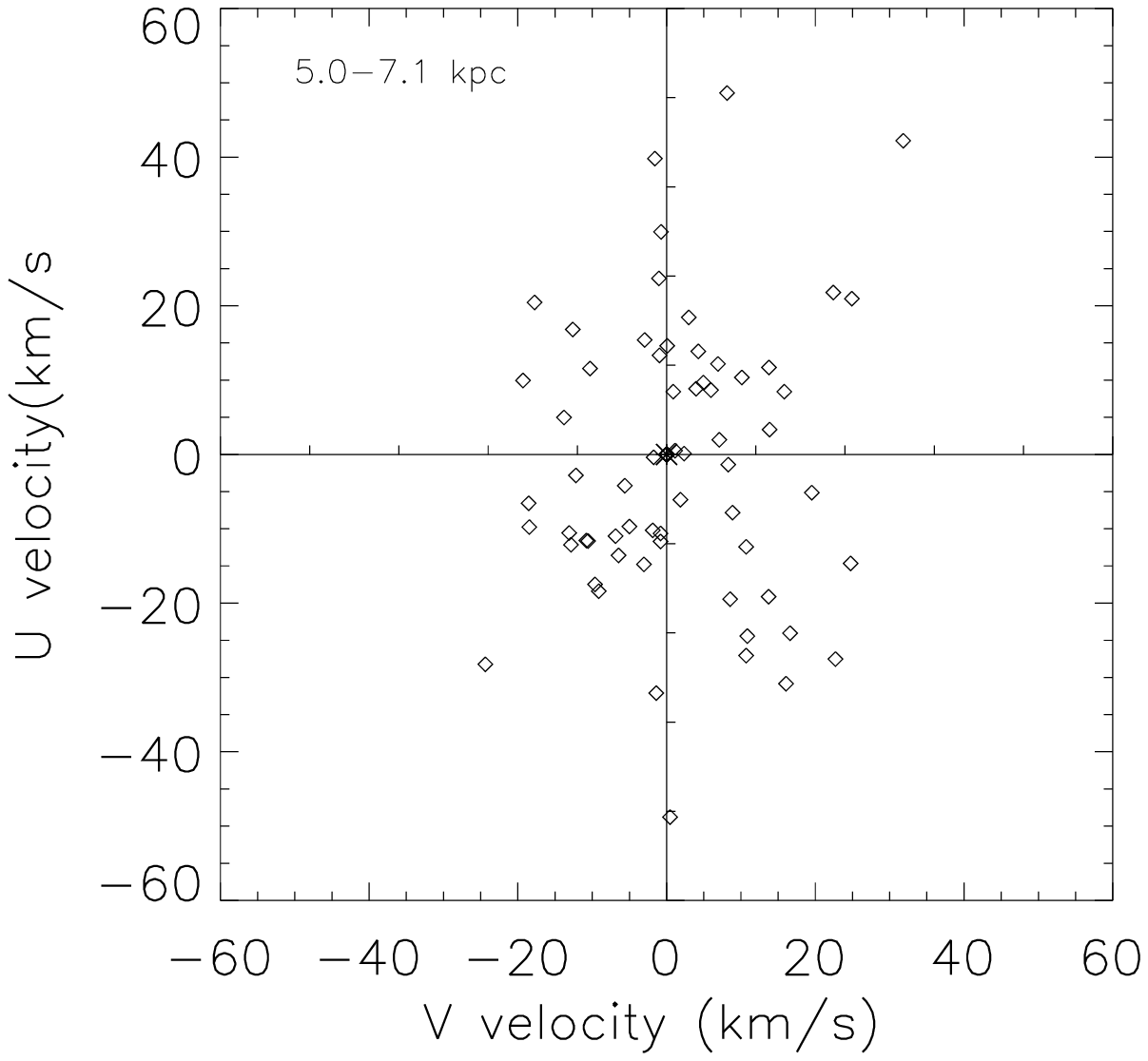}
\caption{Distribution of initial velocities of clusters in the galactic radius
range 5.0  to 7.1 kpc}
\label{figure12}
\end{figure}

\begin{figure}
\plotone{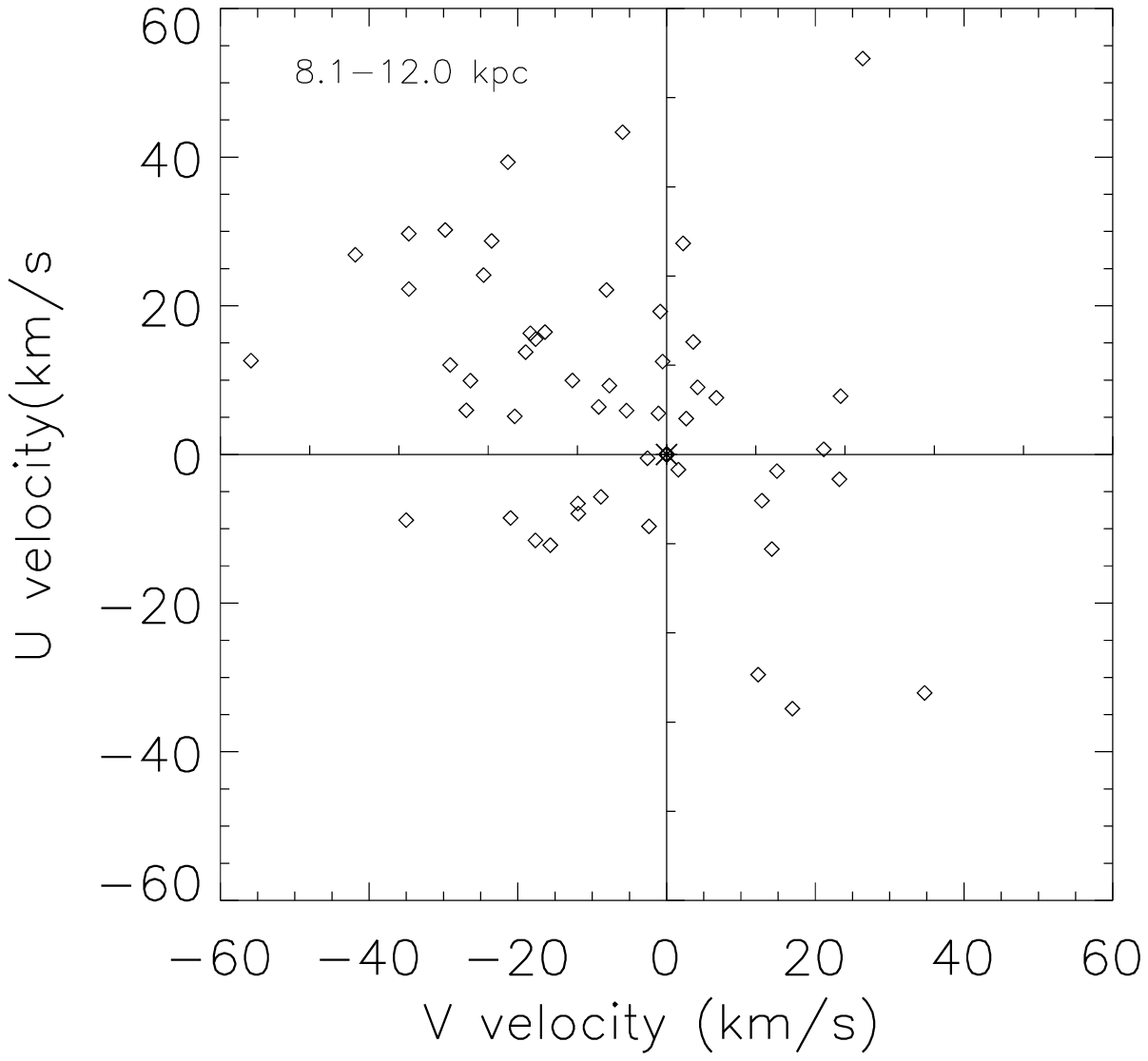}
\caption{Distribution of initial velocities of clusters in the galactic radius
range 8.1  to 12.0 kpc}
\label{figure13}
\end{figure}

\section{Conclusions}

The study of samples of open clusters situated in  small intervals of galactic
radius or rings with width $<$ 1 kpc, reveals the presence of several sequences
of 4 to 12 clusters that have been formed with a same initial direction of the perturbation
velocity, as measured with respect to the direction of circular rotation. The expected
rotation of the perturbation velocity vector from the epicycle theory can be observed
in plots of the angle versus age of the clusters.  In such plots, the sequences of 
clusters along lines that correspond to a same initial angle span ranges of ages
of about 40 to 150 Myr. In a given ring, usually 2, 3 or more such preferential
directions are observed. Probably, the initial angles are dictated by the local
hydrodynamics in the spiral arms. The observations suggest that portions of the 
shock waves associated with the arms are coherent, or keep the same characteristics,
during  period of times of more than 100 Myr. Therefore, the observations
provide an argument in favor of star formation  triggered directly by the 
spiral shocks, and not by secondary processes like the explosion of supernovae. 

If we average the initial direction of the perturbation velocity over a relatively 
large range of galactic radius (a few kpc), the individual coherent
groups become less prominent, or average out, but a broad general distribution
of initial directions appears, with a peak at about -70$^o$, for galactic radii
smaller than R$_0$ (or smaller than the corotation radius). This particular initial 
direction helps the arms to keep their sharply defined aspect, since the clusters do
not move out from the arms before the bright short-lived stars reach the end of their life. 

The property of the cluster formation mechanism  of producing initial perturbation
velocities in a given direction for groups of clusters spanning a range of ages
allows us to perform a direct determination of the epicycle frequency $\kappa$ in
the galactic disk. This measurement shows only a minor dependence on the rotation
curve which is used to perform the corrections for differential rotation, to find
the perturbation velocity. For a given rotation curve, there are two ways of
obtaining the epicycle frequency at a radius $R$. One is to make direct use of the 
expression for $\kappa$ as a function of $\Omega$ and its derivative, and the other
uses the plots of the angle  versus age of the open clusters.  We compared the
results obtained using different rotation curves, based on different values of
$R_0$ and of $V_0$. If we fix for instance $R_0$ and vary $V_0$, we can  find a
 best value of $V_0$, for which the "observed" and the "theoretical" values of 
$\kappa$ coincide.

If $R_0$ = 7.5 kpc is adopted, using the sample of clusters in the range 
$ 7.1 < R < 7.9$ kpc, we find $V_0$= 226$\pm$15 kms$^{-1}$. The same value
 of $V_0$ is obtained when we adopt $R_0$ = 7.0 kpc or $R_0$ = 8.0 kpc. 
 Therefore,  $V_0$= 226$\pm$15 kms$^{-1}$ is a robust result, which does not
 depend  on the precise choice of $R_0$, in the 7-8 kpc range.

We examined samples situated in rings with different galactic radii; for instance with
mean radius between 6 and 8  kpc. The rotation curve $R_0$ = 7.5 kpc , 
 $V_0$= 226 kms$^{-1}$, with a minimum at 7.8 kpc is found to be a
satisfactory one, since it predicts values of $\kappa$ that are consistent with
the observed ones, not only at $R_0$, but also at these different galactic radii.
The ratio $V_0 / R_0$ that we recommend is 30.1 kms$^{-1}$kpc$^{-1}$, which is
larger than that "classical" value obtained from the Oor's constants (A-B),
25 kms$^{-1}$kpc$^{-1}$. However, as we discussed in a previous section, there are
 many recent measurements that result in $V_0 / R_0$ close to 30 kms$^{-1}$kpc$^{-1}$,
including a method that does not depend on local irregularities of the rotation 
curve. The method to determine $V_0$ that we  proposed is a new one, which is something
much needed in view of the conflicting results that have been published in the 
last decade, almost always based on analyses of Oort's constants. 

\acknowledgments{The work was supported in part by the Sao Paulo State
agency FAPESP (fellowship 03/12813-4)by the Conselho Nacional de 
Desenvolvimento Cientifico e Tecnol\'ogico.} 
 
\section{Appendix A}

The present paper deals with open clusters that are spread over distances
distances of several kpc, so that it is convenient to use an individual
frame of reference for each cluster, to describe their motion.
The local frame is chosen to rotate around the galactic center with the
velocity of the rotation curve at the radius where the cluster is at present epoch.
In such a reference frame, in which $\xi_0$ = 0 and $\eta_0$ =0, the MOT's equations
of motion (see section 3.1.2) reduce to: 

$$ \xi = \frac{\eta_0'}{2B}  + \frac{\eta_0'}{2B} cos(\kappa t)
 + \frac{\xi_0'}{\kappa}sin(\kappa t)$$

$$ \eta = \frac{\xi_0'}{2B}  - \frac{\xi_0'}{2B} cos(\kappa t) + \eta_0'\frac{A}{B}t$$
 $$  + \eta_0'\frac{A-B}{\kappa B}sin(\kappa t) $$
  
  where A and B are the Oort's constants. Taking the derivative:
  
  $$ \xi' = \frac{\eta_0'\kappa}{2B} sin(\kappa t)
 + \xi_0' cos(\kappa t)$$
 
 $$ \eta' = \frac{\xi_0' \kappa}{2B} sin(\kappa t)
  + \eta_0'\frac{A}{B} - \eta_0'\frac{A-B}{ B}cos(\kappa t) $$

  The only non-harmonic term is $\eta_0'\frac{A}{B}$, a term that does not
  increase with time and represents a  shift in velocity, which corresponds to the 
  change from the local frame to the frame of the guiding center (the frame in 
  which the motion is described by pure harmonic functions in both axes). At the
  instant t=0 the velocity in the local frame is ($\xi_0', \eta_0'$) and in the 
  frame of the guiding center ($\xi_0', \eta_0'[1-\frac{A}{B}]$).
  
  The initial angle $\gamma$ in the frame of reference of the guiding center
  is:
  
  $$\gamma_{0 GC} = atan(\frac{\xi_0'}{\eta_0'[1-\frac{A}{B}]})  $$ 
  so that:
  $$\delta \gamma = atan (C\times tan [\gamma_0]) - \gamma_0$$
  
  where $\delta \gamma$ is the correction to $\gamma_0$ needed to convert it from the
  local reference frame to that of the guiding center. The constant $C$ obtained
  using the classical values of the constants A =15 kms$^{-1}$kpc$^{-1} $ and B=
  -11 kms$^{-1}$kpc$^{-1} $ (Mihalas \& Binney, 1981, chapter 8) is 0.42; 
  using the Oort's constants adopted by MOT, A= 12 kms$^{-1}$kpc$^{-1} $ and B=
   -15 kms$^{-1}$kpc$^{-1} $, $C$= 0.55. since only an order of magnitude is needed,
   we can adopt $C$= 0.5. The function $\delta \gamma$ is illustrated in Figure A1;
  its average is zero, its maximum about 24$^o$ and rms value 15$^o$ in the worst case
  ($C$= 0.42).

\begin{figure}
\plotone{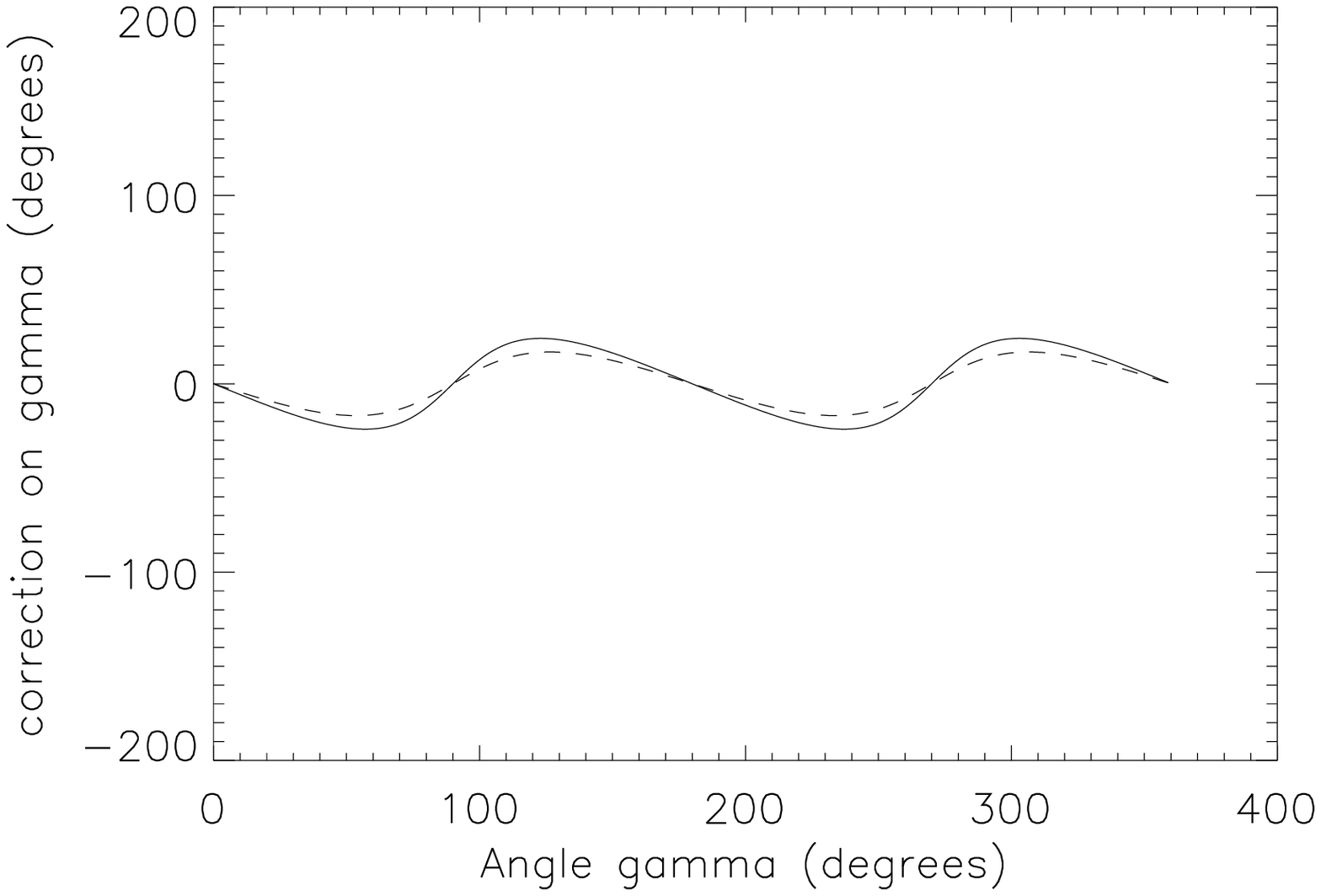}
\caption{Correction to the angle $\gamma$ to move it to the reference frame of
the guiding center of the epicycle motion, as a funtion of $\gamma$. The dashed line 
is with $C$=0.55, full line with $C$=0.42 (see Appendix).}
\label{figA1}
\end{figure}

% -------------------------- Bibliography --------------------------------

\end{document}